# Industry Perception of Security Challenges with Identity Access Management Solutions

Abhishek Pratap Singh
*Department of Informatics*
*King's College London*
London, UK
abhishek.p.singh@kcl.ac.uk

Ievgeniia Kuzminykh
*Department of Informatics*
*King's College London*
London, UK
*Department of Infocommunication Engineering*
*Kharkov National University of Radio Electronics*, Ukraine
ievgeniia.kuzminykh@kcl.ac.uk

Bogdan Ghita
*School of Engineering, Computing and Mathematics*
*University of Plymouth*
Plymouth, UK
bogdan.ghita@plymouth.ac.uk

*Abstract*— Identity Access Management (IAM) is an area posing significant challenges, particularly in the context of remote connectivity and distributed or cloud-based systems. A wide range of technical solutions have been proposed by prior research, but the integration of these solutions in the commercial sector represent steps that significantly hamper their acceptance. The study aims to outline the current perception and security issues associated with IAMs solutions from the perspective of the beneficiaries. The analysis relies on a series of interviews with 45 cyber security professionals from different organisations all over the world. As results showed, cloud IAM solutions and on premises IAM solutions are affected by different issues. The main challenges for cloud based IAM solutions were Default configurations, Poor management of Non-Human Identities such as Service accounts, Poor certificate management, Poor API configuration and limited Log analysis. In contrast, the challenges for on premise solutions were Multi Factor Authentication, insecure Default configurations, Lack of skillsets required to manage IAM solution securely, Poor password policies, Unpatched vulnerabilities, and compromise of Single-Sign on leading to compromise of multiple entities. The study also determined that, regardless the evolving functionality of cloud based IAM solutions, 41% of respondents believe that the on premise solutions more secure than the cloud-based ones. As pointed out by the respondents, cloud IAM may potentially expose organisations to a wider range of vulnerabilities due to the complexity of the underlying solutions, challenges with managing permissions, and compliance to dynamic IAM policies.

*Keywords— Identity Access Management, IAM, On Premise, Cloud, Access Control, Vulnerability, Authentication*

## I. INTRODUCTION

Identity and Access Management (IAM) is pivotal in safeguarding digital and at times physical identities. IAM solution is being extensively used in a typical on-premises environment where deploying organisations own everything including user administration, cost of software, underlying support infrastructure along with security aspects such as vulnerability and patch management, as well as penetration testing. On the other side, in a cloud-based model such as Infrastructure as a server (IaaS), Software as a Service (SaaS) IAM is addressed by the Cloud service provider (CSP).

In the realm of cybersecurity, IAM stands as a critical cornerstone. Simultaneously, cloud computing has emerged as a transformative model for delivering IT services via the Internet, offering scalability, flexibility, and cost efficiency. However, this paradigm shift brings various security challenges, particularly within IAM.

A study [1] shows that Cloud IAM solutions occupy 42% of the IAM market with expected growth of 18.3% between 2022 and 2032, and sales of cloud IAM are expected to reach US$ 25,539.2 million by 2032.

Checkpoint cloud security report 2023 [2] indicates that around 24% organisations have faced public cloud related incidents. The previous Checkpoint report in 2022 [3] showed that 15% of all cloud incidents were linked to compromising of IAM; report also states that 72% of respondents are using Microsoft Azure, 69% are using Amazon Web Services (AWS) and 34% are using Google Cloud Platform (GCP) as their IaaS providers, yet 54% of this audience conceded to the fact that for Cloud security they rely on independent security service provider. Despite highlighting the overall concerning picture, neither of the two reports details specific incidents or concerns related to the cloud based IAMs.

Given the increasingly distributed range of technical solutions for IAM, coupled with the seamless interconnectivity of systems in such environments, organisations are (justifiably) concerned when required to deploy more flexible IAM, potentially as a cloud service rather than a controlled on-premises solution. Therefore, the aim of this study is to explore the security concerns and weaknesses of cloud and on premise solutions for access management from the perspective of their beneficiaries – the companies requiring such implementations.

## II. BACKGROUND AND RELATED WORKS

In order to understand the current level of threat posed by the IAM vulnerabilities and weaknesses, this section discusses a series of specific security breaches, based on their public reporting.

A study by the monitoring company ManageEngine [4], which focused on cyber incidents linked due to IAM solutions failure over last decade, described a breach of the Deloitte global email server infrastructure via an administrator account which was protected by a single password without any multi-factor authentication (MFA). In a similar case, the online shopping giant eBay and the major home improvement retailer HomeDepot became victims of data exfiltration as credential of small group of employees were compromised. According to the Checkpoint reports [2, 3] the account compromise composed 29% in 2022 and 16% in 2023 of all security incidents related to public cloud.

The study of Gofman and Dahan [5] found common weaknesses in the IAM model of top three players of cloud industry AWS, Google Cloud Platform (GCP) and Azure linked to dangerous permissions categorised under Assignment, Code Execution, Grants and Delegation and New Credentials.

The study [6] discussed the authentication challenges with cloud based IAMs related to MFA and single sign-on (SSO). MFA is a powerful defence against various authentication attacks, including phishing and password breaches. However, implementing and managing MFA in cloud environments can be challenging for a number of reasons, such as the disparity in MFA support across different cloud providers. Some providers may not have a complete MFA offer, making it challenging for organisations to choose and enforce MFA consistently, especially for users utilizing diverse devices and applications.

The authors of [7] discussed the challenges related to the policies and rules that dictate user permissions to cloud resources, a fundamental aspect of IAM. IAM policies are an essential security measure, as they help ensure users only have access to the resources they need. However, IAM policies can also be challenging to create, manage and understand.

While there are many studies that investigate the security of IAMs separately, on premises [8, 9] or in the cloud [10-12], existing literature does not provide a comparative analysis of IAM deployment models. An observational analysis of existing studies was performed in [13] to compare two models in terms of cost effectiveness, ease of management, scalability and agility, constant updates, and compliance. Another study [14] as based on the surveys of professionals to get insights on usage of IAM solutions in the companies. While the study tackled outsourcing of IAM vs Internal management, size of IAM team and compliance with zero trust principles, it had only 3 questions out of 8 related to IAM, which mostly focused on collecting statistical data.

It is apparent that existing studies tend to investigate IAM from a model, conceptual perspective. In contrast, this study aims to gather a deeper insight into IAMs through the professional experience on the usage of IAMs by companies and to highlight its potential vulnerabilities and security challenges based on its actual use in the industry.

### III. METHODOLOGY

#### A. Research Questions

Based on the aim of this study, we pose the following research questions:

*RQ1: What are the key challenges and vulnerabilities associated with cloud based IAMs?*

*RQ2: On-Premises versus Cloud, which IAM model is more secure?*

#### B. Data Collection

In order to collect the data, we interviewed cyber security professionals of leading organisations to learn from their day-to-day experience about security weaknesses and challenges witnessed during different phases of deployment of on-premises and cloud-based IAM solutions. The distribution of respondents in terms of expertise and industry sector is shown in Fig.1; the majority of respondents are holding the position of Chief Information Security Officer (CISO) and IT Security Administrator, among the rest are Chief Technology Officer (CTO) and Directors.

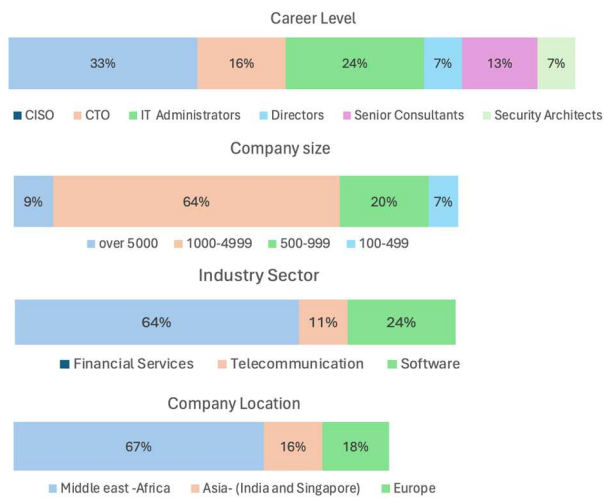

Fig. 1. Respondents data

The following set of questions were asked to respondents:

- What factors influenced your organisation's decision to choose between on-premised and cloud-based IAM solutions, and how do these decisions align with your security strategy?
- What motivated your organisation's shift to cloud based IAM solutions?
- In your experience which IAM deployment model is more secure?
- As CISO, what key security challenges have you faced with IAM solutions be it on premised or cloud based?
- How has cloud based IAM impacted your organisation's security? Have you suffered a breach in recent past?
- How does your organisation ensure compliance with regulations while using cloud based IAM, and what challenges have you encountered?
- Does your organisation deal with multiple cloud service providers?

The participant responses were further analysed using the Lexalytics content analysis tool to get a quantitative view of subject in scope. The interviews and surveys were conducted in the period of November-December 2023. The participants were provided with written assurance that their names and affiliation will not be disclosed at any point in time.

## IV. RESULTS AND ANALYSIS

The survey identified top six challenges related to on-premises IAM deployment model illustrated on Fig.2. The challenges reported by interviewees included not having the MFA, the prevalence of default configurations, the workforce missing the skillsets required to manage IAM solution securely, and the increasing reliance of Single-Sign on authentication, leading to exposure of multiple entities when accounts are compromised..

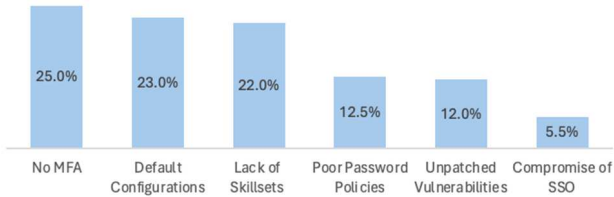

Fig. 2. Challenges with on-premise IAM solutions.

For cloud based IAM solutions, Default configurations, Poor management of Non- Human Identities such as Service accounts, SaaS applications, services and APIs, Poor certificate management, Poor access review, Poor API configuration and limited Log analysis with multi-cloud were among top challenges, as illustrated on Fig.3.

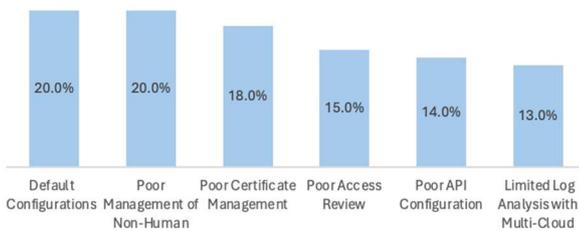

Fig. 3. Key weaknesses with cloud based IAMs.

The analysis revealed that 76% of organisations use more than two cloud service providers (CSP). The survey results showed that 53.4% of the security professionals believe that cloud based IAM solutions are more secure versus 41% who prefer the security of on premises IAM solutions (Fig.4). The high percentage of on-premises solutions supporters could be explained by the growing threat landscape due to the integration of IAM into cloud [15, 16]. In addition, given that many organisations use several CSPs, this requires an IAM solution supporting multiple cloud environments, creating a local identity for each of the application that makes it very difficult to keep track.

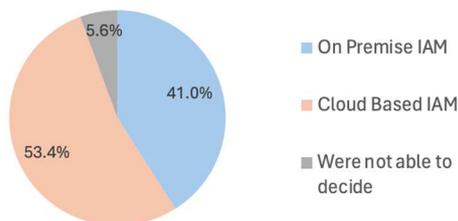

Fig. 4. Security comparison of on premise and cloud based IAM solutions.

In addition, to the respondents preferences, the analysis also highlighted the risks associated with cloud-based IAM vulnerabilities, outlined in the remainder of this section.

### A. Data Breaches and Unauthorized Access

One of the pivotal outcomes of IAM vulnerabilities resides in the data breaches. Data breaches involve unauthorized attackers getting access to confidential information [17]. These incidents can result in the disclosure or customer data theft, substantial financial setbacks, and significant damage to the organisation's reputation. Any unauthorised access into cloud resources gives attackers the opportunity to to steal data, disrupt operations, or launch an attack on other systems.

A data breach can result in significant financial losses and can potentially result in fines and penalties from regulators. For instance, the European Union's General Data Protection Regulation (GDPR) empowers government agencies to impose fines of up to 4% of a company's global annual turnover or €20 million, against organisation that fails to fulfil the obligations to protect the personal data.

### B. Financial Losses

IAM vulnerabilities, when exploited, can lead to significant financial losses [18]. Beyond the direct costs associated with addressing the security incident, organisations must consider the expenses related to investigating the incident, implementing security improvements, and notifying affected customers. In addition, the organisations may face indirect financial consequences, such as decreased revenue and increased insurance premiums. Customers may lose confidence in a company, leading to reduced business, lost sales, and long-term financial impact.

### C. Reputational Damage

IAM vulnerabilities and subsequent data breaches can cause serious reputational damage to an organisation. Customer trust is of utmost importance in business, but it is a rather volatile metric, which is significantly impacted by data breach incidents. Customers may view such organisations as unreliable or negligent, which can result in a loss of business, decreased market share, and a damaged brand image. Media coverage of data breaches can exacerbate reputational damage: negative headlines and news stories can spread quickly, reaching a broad audience, and further tarnishing an organisation's reputation. Restoring trust and credibility can be a long and challenging process, often requiring significant investments in public relations and marketing efforts.

### D. Compliance and Regulatory Issues

The participants highlighted that the organisations are subject to strict regulatory obligations with regard to data security. The consequences of failure to comply with these mandates could be extremely serious. The organisations that fail to comply regulatory benchmarks may find their ability to engage with specific government entities or other organisations severely limited, thus, in turn, limiting their market influence and potential prospects.

Analysing the potential impact of these IAM vulnerabilities using the security risk management methods [19] highlights the importance of implementing robust security protocols and best practices.

## V. DISCUSSION

From the analysis conducted in the previous sections, we summarize the following findings with regards to the research questions.

*RQ1: What are the key challenges and vulnerabilities associated with cloud based IAMs?*

The results of the survey showed that the IAM challenges identified by the security professionals for on premise and cloud-based solutions differ significantly. The on premise IAM deployment scenarios tend to lack the skilled personal required to deploy and manage the hardware, software, and associated tools. Interestingly, the respondents in our study did not identify this as challenge for cloud solutions, in contrast to the survey from Checkpoint which states that the lack of skill is the biggest challenge (58%) [2, 3]. It is obvious that there is an economical challenge associated with the cost of deployment and the human resource required to manage the IAM, but our results showed that, once the IAM solution is deployed, other security challenges come onto the first place. All the challenges highlighted in the survey relate to technical or configuration difficulties and can lead to potential vulnerabilities. Default configuration, SSL certificate management, API between company side and cloud solution, access method using MFA, and SSO were reported as the areas of most concern for organisations with IAM solutions.

The research study showed that organisations encounter not only technical issues with IAM within cloud environments, but also suffer from vulnerabilities related to unauthorised access, reputational damage, and compliance with regulations. Cloud service providers offer an extensive spectrum of permissions, thereby putting organisations in the position to deal with the implementation of various authentication and authorisation protocols to access the cloud which might create vulnerabilities related to misconfiguration of such protocol and requires additional work resources. The extensive capabilities of cloud settings offer many options, creating additional layer of complexity to IAM implementation. Moreover, the task of managing the permissions becomes a huge barrier. Organisations are therefore faced with the challenges of managing multiple credentials required for various cloud services which have been already flagged by previous studies [15 20, 21]. Dynamic IAM policies represent a persistent challenge, as the ever-evolving nature of cloud environments requires constant monitoring and adjustments to keep IAM policies and configurations up to date.

*RQ2: On-Premises versus Cloud, which IAM model is more secure?*

Determining a concrete answer to RQ2 is difficult based on the limited sample size during survey. The current opinions were nearly equally split between on-premises (41%) and cloud solutions (53%). This can be explained by the growing security threat landscape when moving to cloud environment and associated complexity of managing the multiple cloud platforms, applications and accounts. According to a Forrester report [22], only 12% of organisations are fully relying on cloud-based identity management.

## VI. CONCLUSIONS

In summary, this study has highlighted the significant challenges and vulnerabilities related to IAM both on premise and in cloud settings. The research identified complexities in cloud permission models given the broader threat landscape, the necessity for various methods of authentication and authorization, challenges in managing credentials, and the dynamic nature of IAM policies as primary obstacles for wider adoption of cloud based IAM solutions.